# Time and Knowability in Evolutionary Processes

Elliott Sober* and Mike Steel†

*Philosophy Department, University of Wisconsin, Madison, WI, USA
†Biomathematics Research Centre, University of Canterbury, Christchurch, New Zealand

*Abstract*: Historical sciences like evolutionary biology reconstruct past events by using the traces that the past has bequeathed to the present. The Markov Chain Convergence Theorem and the Data Processing Inequality describe how the mutual information between present and past is affected by how much time there is in between. These two results are very general; they pertain to any process, not just to the biological processes that occur in evolution. To study the specifically biological question of how the present state of a lineage provides information about its evolutionary past, we use a Moran process framework and consider how the kind of evolutionary process (drift, and selection of various kinds) at work in a lineage affects the epistemological relation of present to past.

What is the epistemic relation of present to past? Absent a time machine, we are trapped in the present and must rely on present traces to learn about the past. There are memory traces inside the skull, but outside there are tree rings, fossils, and traces of other kinds. People use these traces to reconstruct the past. Sometimes they simply assume that the traces provide unerring information about the past, but often they realize that the jump from present to past is subject to error. A bevy of epistemic concepts can be pressed into service to investigate the relation of present traces to past events, ranging from strong concepts like knowledge and certainty to more modest ones like justified belief and evidence.

We are interested in how the natural processes connecting past to present constrain our ability to know about the past by looking at the traces found in the present. An optimistic view of these processes is that the past is potentially an open book; all we need do is understand the connecting processes correctly and look around for the right traces. If the relation of the past state of a system to its present state were deterministic and one-to-one, this optimistic view would be correct. If only we could know the present state with sufficient precision, and if only we could grasp the true mapping function that connects present and past, we would be home free. This optimism is something that Laplace (1814, p. 4) affirmed when he discussed *une intelligence* (now referred to as "a demon"):

> We may regard the present state of the universe as the effect of its past and the cause of its future. An intellect which at a certain moment would know all forces that set nature in motion, and all positions of all items of which nature is composed, if this intellect were also vast enough to submit these data to analysis, it would embrace in a single formula the movements of the greatest bodies of the universe and those of the tiniest atom; for such an intellect nothing would be uncertain and the future just like the past would be present before its eyes.

It is worth noting that determinism is not sufficient for the optimistic view to be true; without the one-to-one assumption, distinct states of the past may map onto the same state of the present, with the consequence that the exact state of the past cannot be retrieved from even a perfectly precise grasp of the present state.

Is determinism necessary for the optimistic view to be true? It is not, provided that we set to one side strong concepts like knowledge and certainty and take up an epistemic evaluation that is more modest. Consider, for example, a process in which the system is, at each moment, in one of two states (coded 0 and 1). Suppose Past=0 makes Present=0 extremely probable (say, 0.96) and that Past=1 makes Present=1 extremely probable as well (say, 0.98). This means that when we observe the system's present state, we gain strong evidence that discriminates between the two hypotheses Past=0 and Past=1. We can't infer from Present=0 that the past state was certainly 0; in fact, we can't even infer that the past state was probably 0. But we can conclude that the observation favors the hypothesis that Past=0 over the hypothesis that Past=1. This conclusion is licensed by what Hacking (1965) calls the Law of Likelihood:

> Observation O favors hypothesis $H_1$ over hypothesis $H_2$ if and only if $\Pr(O \mid H_1) > \Pr(O \mid H_2)$.

Royall (2007) suggests that this qualitative principle should be supplemented by a quantitative measure of favoring:

The degree to which O favors $H_1$ over $H_2$ is given by the likelihood ratio $\frac{\Pr(O|H_1)}{\Pr(O|H_2)}$.

Royall further suggests that a reasonable convention for separating strong evidence from weak is a ratio of 8. If we adopt Royall's suggestion, we must conclude that the probabilistic process just described entails that Present=0 provides strong evidence favoring Past=0 over Past=1, since the likelihood ratio is $\frac{0.96}{0.02} = 48$.

This simple example should not be over-interpreted. *If* there were a process connecting past to present in the way described, the present would provide strong evidence about the past. Do not forget the *if*. Perhaps there are such processes, especially when the past we are talking about is the recent past. But what if we consider not just the recent past, but past times that are more and more ancient? How does increasing the temporal separation between present and past affect the amount of information that the present provides about the past?

**Two classic theorems**

The Markov Chain Convergence Theorem provides an answer to the question just posed (Cover and Thomas 2006). Consider a system that at any time is in one of *n* possible states ($s_1$, $s_2$, ..., $s_n$). For simplicity we'll think of the system as evolving in discrete time steps. We stipulate that the system has the following four properties:

(*constant transition probabilities*) There is a fixed set of transition probabilities; these describe the system's probability of being in state $s_i$ at time *t*+1, given that it was in state $s_j$ at time *t* (for all *i*,*j*). These transition probabilities are constants; they do not change values as the system evolves.

(*irreducibility*) For any two states, each is accessible from the other, in the sense that there is a chain that goes from each eventually to the other where each change in the chain has nonzero probability. It needn't be possible to move directly from each state to every other in a single jump.

(*aperiodicity*) This is delicate to state in full generality; see Häggström (2002) for details. For present purposes, we state a sufficient condition: For any state that the system is in, it's possible that it remains in that state at the next time step.

(*the Markov property*) For any two times $t_1 < t_2$, the state of the system at $t_1$ screens-off the system's history prior to $t_1$ from the state at $t_2$. That is,

Pr(system is in state *y* at $t_2$ | system is state *x* at $t_1$) =
Pr(system is in state *y* at $t_2$ | system is state *x* at $t_1$ & system's history prior to $t_1$), for all states *x* and *y*.

For any system of this sort, the following result holds:

> *The Markov Chain Convergence Theorem*: If a system has constant transition probabilities, and the system is irreducible, aperiodic, and satisfies the Markov property, then I(Past, Present) approaches zero as the time separating past and present approaches infinity.

I(X;Y) is the "mutual information" linking the two variables. If the variables are discrete, the formula for this quantity is:

$$I(X;Y) = \sum_{y \in Y} \sum_{x \in X} p(x,y) \log\left(\frac{p(x,y)}{p(x)p(y)}\right)$$

where *p(x,y)* is the joint probability that *X = x* and *Y = y*, and *p(x)*, *p(y)* are the (marginal) probabilities that *X = x*, and that *Y = y*, respectively. Mutual information measures how much (on average) you learn about the state of one of the variables by observing the state of the other. Its value is zero when *X* and *Y* are independent; otherwise it is positive. Mutual information is symmetrical: I(X;Y) = I(Y;X). The decay of information described by the Markov chain convergence theorem occurs exponentially fast (Sober and Steel 2011, Proposition 6).

The convergence theorem does not insure a monotonic decline in information as the temporal separation of past from present is increased. That extra element is provided by a different result, the so-called "Data Processing Inequality" (Cover and Thomas 2006):

*The Data Processing Inequality*: In a causal chain from a distal cause D to a proximate cause P to an effect E, if P screens-off D from E, then I(E;D) is less than or equal to both I(E;P) and I(P;D).

For a discrete-state process, these two inequalities are strict whenever P is neither perfectly correlated with D or E nor independent of them (See Appendix A). The data processing inequality does not require that the process linking D to P is the same as the process linking P to E. That is an additional difference between it and the Markov Chain Convergence Theorem.

The information processing inequality is "chain internal." It does not say that the present always provides more information about recent events than about ones that are older. Consider Figure 1. Suppose that R screens-off A from $D_1$&$D_2$. Then the information processing inequality says that $I(D_1$&$D_2; R) \geq I(D_1$&$D_2;A)$. It does not say that $I(D_1$&$D_2$&…&$D_{100}; R) \geq I(D_1$&$D_1$&…&$D_{100}; A)$. It is perfectly possible that the hundred descendants that now exist provide more information about A than they do about R. Note that R does not screen-off A from $D_1$&$D_2$&…&$D_{100}$.

Figure 1: A case in which it is possible for the present to provide more information about an ancient event (A) than about a more recent event (R).

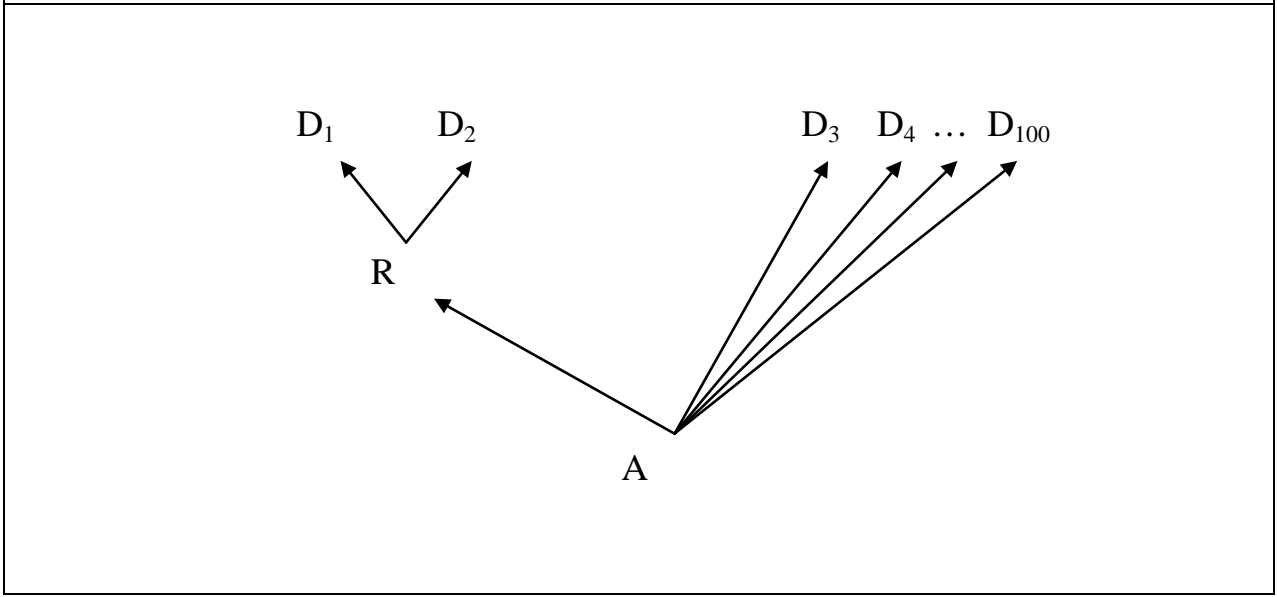

Both the Markov Chain Convergence Theorem and the data processing inequality are very general. They characterize any system whose laws of motion have the requisite probabilistic features. The system might be a chamber of gas, but it also might be an evolving population of organisms. Indeed, if there were disembodied spirits that changed probabilistically, the results would apply to them. Both results are *more general than physics* – they cover the systems and properties that are discussed in the laws of physics, but they also apply to systems and properties that are not. In addition, both results are *a priori* mathematical truths, though of course it is an empirical matter whether a given system satisfies the antecedent of the conditional that each result expresses (Sober 2011).

**Five Evolutionary Processes**

Since our interest here is in how evolutionary processes affect the amount of information that the present provides about the past, it is worth making clear how the Markov Chain Convergence Theorem applies to models of biological evolution. Moran (1962) models of evolution will be our work horse in what follows. We consider a population containing $N$ individuals. The population evolves through a sequence of discrete temporal "moments" (how long a moment is won't matter). At each moment, one of those $N$ individuals produces a copy of itself and one of those $N$ individuals dies. We consider two traits A and B; each individual has one of them or the other. At any moment, the population is in one of $N+1$ states (ranging from 0% A to 100% A). This Moran framework can be articulated in different ways to represent different evolutionary processes. For example, if individuals are chosen at random to reproduce and die, then we have a drift process. Selection processes of different kinds can be represented by letting A individuals have chances of dying or reproducing that differ from those possessed by B individuals. A population undergoing a Moran process forms a Markov chain, with its recent past screening-off its more remote past from the present.

Suppose we observe the population in the present and see that all $N$ individuals are in state A. How much information does that observation provide about the state of the population at some earlier time? If all states are accessible to each other (which requires that mutations can prevent the population from getting "stuck" at 100% A or 100% B), then the Markov Chain Convergence Theorem applies and so the mutual information asymptotes to zero with time. However, if there is no mutation, then the population will evolve to either 100% A or 100% B and will stay there. In this case, the present state of the population provides information about its past even if the two are infinitely separated. For example, if we observe that the population is now 100% A and the population has been evolving by drift, this observation favors the hypothesis that the population was 100% A at some earlier time over the hypothesis that it was 5% A, even if present and past are infinitely separated.

John Maynard Keynes (1924, ch. 3) once said that "in the long run, we are all dead." His point was to pooh-pooh the relevance of claims about the infinite long run. What should matter to us mortals is finite time. This point applies to the bearing of the Markov Chain Convergence Theorem on our knowledge of the evolutionary past. Who cares if mutual information goes to zero as the time separating present from past goes to infinity? Life on earth is a mere 3.8 billion years old. What is relevant is that information decays monotonically in Markov chains. But in addition, we know that there are different kinds of evolutionary process. Which of these speeds the loss of information and which slows it?

The five processes we want to investigate are represented in Figure 2; each of them can result in our present observation that all *N* of the individuals in the population now have trait A. In (i), trait A reached 100% representation because it was favored by selection. In (iii), 100% A evolved in spite of the fact that there was selection against it. In (ii) the traits are equal in fitness, so the population drifted to 100% A. In (iv), selection favored the majority trait. And in (v), selection favored the minority trait. We are not asking which of these process hypotheses is most plausible, given the observed state of the population at present. Rather, we want to explore what happens to information loss under each of these five scenarios, in each case thinking of the process in the context of the Moran framework of a finite population of fixed size *N*.

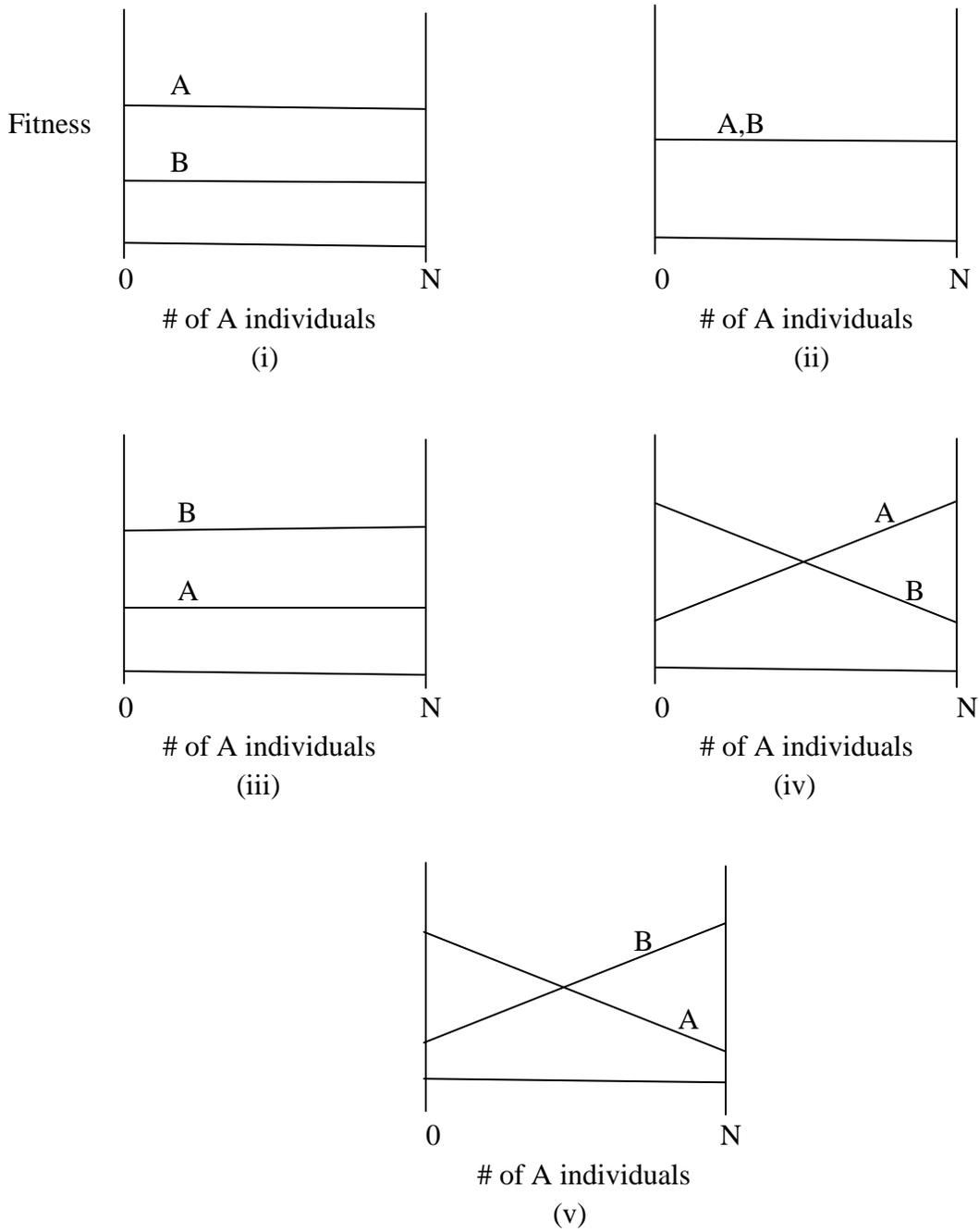

Figure 2: Five processes that can result in all *N* of the individuals now in the population having trait A.

To compare the five processes in this respect, we calculate the following likelihood ratio for each of them:

$$R_{ij} = \frac{\Pr(Present = N \mid Past = i)}{\Pr(Present = N \mid Past = j)}, \text{ for } i>j.$$

The observation is that all $N$ of the individuals now in the population have trait A. Does this observation favor the hypothesis that there were exactly $i$ individuals at some past time who had trait A over the hypothesis that there were exactly $j$, where $i>j$?  It does; $R_{ij} > 1$ for each of the five processes we are considering (see Appendix B).  Our question is how the magnitude of $R_{ij}$ depends on the underlying evolutionary process.

We begin by adopting an assumption that we treated above with Keynesian disdain.  Let's assume that the temporal separation of past and present is infinite and that there is zero mutation.  For each of the processes we're considering, we now can describe what the value of $R_{ij}$ is for each pair of values for $i$ and $j$ such that $i>j$.  For example, under neutral evolution the value of $R_{ij}$ is $\frac{i}{j}$.  The values for the other four processes are given in the Appendix B.  The ordering of the $R_{ij}$ values for the five processes is depicted in Figure 3.

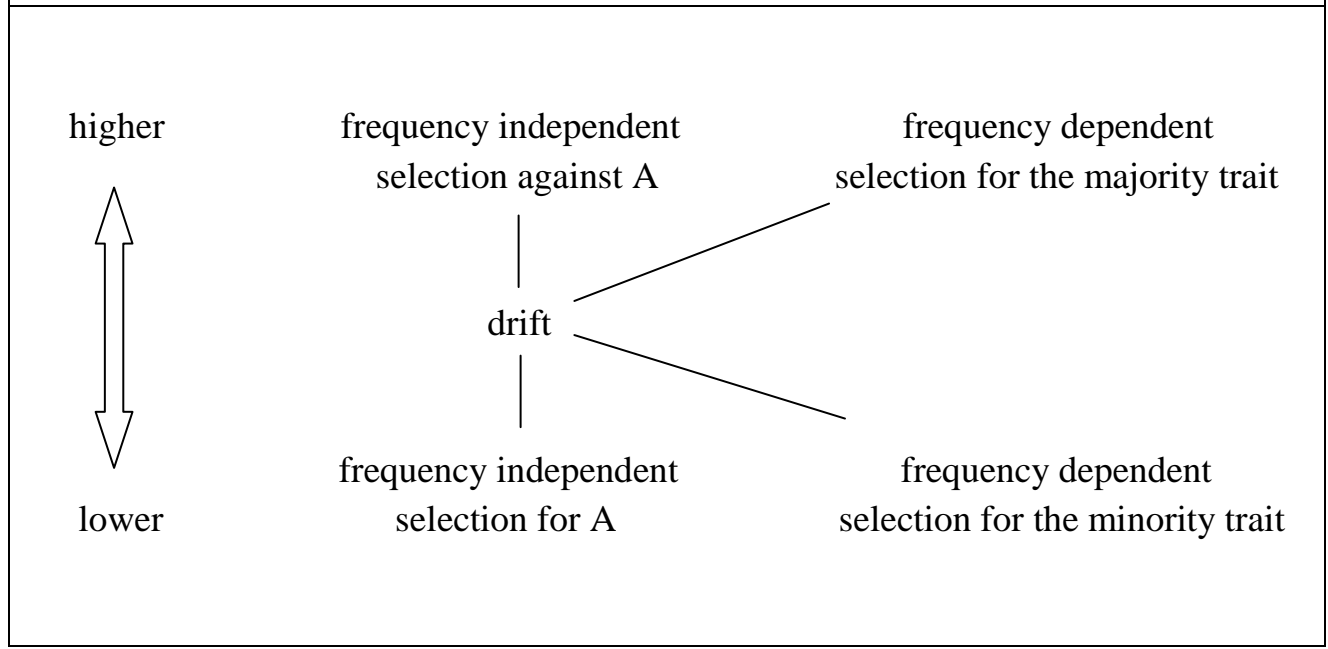

Figure 3: Comparing $R_{ij}$ values for five processes, assuming infinite temporal separation of Past from Present and zero mutation.

higher / lower

frequency independent selection against A

frequency dependent selection for the majority trait

drift

frequency independent selection for A

frequency dependent selection for the minority trait

Let's first consider three of the cases described in Figure 3 -- drift and the two cases of frequency independent selection. The ordering of $R_{ij}$ values for these three processes means that the observation that all the individuals in the population now have trait A provides more information about the past state of the population the less probable it was that A would evolve to fixation. Selection for A is at the bottom of the pile, with neutrality next, and selection against A at the top. This result echoes an insight that Darwin expresses in the *Origin*:

> … adaptive characters, although of the utmost importance to the welfare of the being, are almost valueless to the systematist. For animals belonging to two most distinct lines of descent, may readily become adapted to similar conditions, and thus assume a close external resemblance; but such resemblances will not reveal – will rather tend to conceal their blood-relationship to their proper lines of descent (Darwin 1859, p. 427).

Darwin illustrates this idea by giving an example: whales and fish both have fins, but this isn't strong evidence for their common ancestry, since the trait is an adaptation for swimming through water. Far stronger evidence for common ancestry is provided by similarities that are useless or deleterious. This idea is so important to Darwin's reasoning that one of us has called it "Darwin's Principle" (Sober 2011). Darwin's topic in the passage quoted is inferring common ancestry, not inferring the past state of a lineage from its present state, but the epistemology is the same.

There are two cases of frequency dependent selection represented in Figure 3. One of them (frequency dependent selection for the majority trait) shows that it would be an over-statement to say that the current state of the population (all individuals having trait A) provides little evidence concerning the population's past state if trait A evolved because of natural selection. It matters a great deal what sort of selection process we're talking about. Frequency dependent selection for the majority trait is better than drift as far as the information harvest is concerned. Figure 3 also locates the evidential meaning of frequency dependent selection for the minority trait; it has an informational yield that is worse than that provided by drift. As noted in the Appendix, our results for both cases of frequency dependent selection require the assumption that $j < N/2$.

Figure 3 provides only a partial ordering of the five cases depicted. The reason for this is that a comparison of, say, frequency independent selection against A with frequency dependent selection for the majority trait would depend on the values of specific parameters.

We now can remove the idealization of infinite time and zero mutation. The ordering of the $R_{ij}$'s for the five processes, when time is infinite and mutation is zero, is the same as the ordering of those processes for the following slightly different likelihood ratio, when time is finite (and sufficiently large) and there is a sufficiently small mutational input:

$$\frac{\Pr(Present = N \mid Past \approx N)}{\Pr(Present = N \mid Past \approx 0)}$$

(see Appendix C). Here 'Past ≈ $N$' and 'Past ≈ 0' just mean any state close to $N$, and to 0 (respectively), which are then held fixed across the five models. With mutational input, the Markov Chain Convergence Theorem applies to all five processes. The mutual information between present and past declines monotonically as their temporal separation increases, but the decline is faster under some processes than it is under others.

Our results are derived within the setting of the Moran model of population genetics. This is a finite-state Markov chain that forms a 'continuant process' (Ewens 2010). That is, at each step the number of individuals carrying a particular allele in the population of fixed size $N$ either goes up by 1, or down by 1, or it stays the same. We would expect similar conclusions concerning the partial order of the $R_{ij}$ ratios for other continuant processes, provided the transition probabilities faithfully reflect the various types of selection being compared.

We have emphasized that loss of information within a lineage is a fact of life. However, the branching that takes place in evolution is a force that pushes in the opposite direction, since it creates *new* lineages. As illustrated in Figure 1, it is possible for ancient ancestors to have more present day descendants than more recent ancestors do, and this means that the information lost to the passage of time can be offset by the proliferation of descendants that each bear witness to the ancient ancestor's state. If the process is a symmetric one on two states it is possible to describe precisely how often branching must occur if information loss is to be offset in this way (Evans et al. 2000); for more general processes, we must usually be content with upper and lower bounds.

It might be asked why we need to worry about observing live organisms to infer the characteristics of their ancestors. Doesn't observing fossils provide a simpler and more definitive solution? Our answer has three parts. First, one can't assume that a fossil comes from an ancestor of extant species; it may just be an ancient relative. Second, fossils provide evidence about the morphological hard parts of ancient organisms; molecular characters, not to mention phenotypic features of physiology and behavior, typically do not fossilize. And finally, fossil traces degrade and are subject to the Markov chain convergence theorem if their probabilistic laws of motion take the relevant form.

**The Impact of Branching on Information**

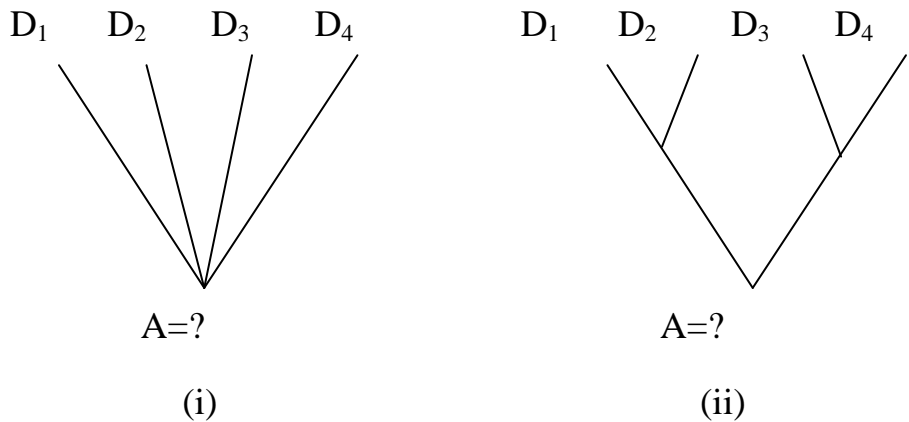

Figure 4: Does observing the four leaves of the star phylogeny (i) provide more information about the state of the root than observing the four leaves of the bifurcating phylogeny (ii)?

Just as the process at work in lineages has an impact on information loss, so too does the topology of the branching process itself. It might seem intuitive to conjecture that the star phylogeny shown in Figure 4i is "better" than the bifurcating topology shown in Figure 4ii in the sense that the former topology allows the observations to provide more information about the root than the latter topology does. In order to hold other factors fixed, we assume that the two topologies have the same number of leaves and that the process at work in branches is the same in the two topologies (in particular, the expected number of substitutions (branch lengths) between the root and each leaf match up for the two trees). The conjecture just stated seems reasonable, since in Figure 4i the observations are independent of each other, conditional on the state of the root, whereas in 4b, the observations are not conditionally independent. The guess is correct for a two-state symmetrical process when we compare the two trees in Figure 4 (Sober 1989, p. 280). More generally, this guess holds whenever we compare any binary tree with a star phylogeny on the same number of leaves when a two-state symmetric process is at work (Evans et al. 2000). But, surprisingly, it isn't true for every symmetric Markov processes on five or more states when the substitution rate lies in a certain region and the two trees have the same (sufficiently large) number of leaves (Sly 2010).

**Conclusion**

In summary, a simplistic view of evolution as an "information-destroying process" overlooks some tantalizing details. For some special processes (e.g. zero mutation in simple population genetic models), information never completely disappears, even after infinite time. At the other extreme there are certain (discrete time) processes for which the information can collapse completely to zero in finite time (Mossel, 1998; Sober and Steel, 2011, p.233). The more usual situation lies between these two extremes; for processes subject to the Markov chain convergence theorem, the information between past and present decays at an exponential rate and vanishes only in the limit. Here we can still compare the relative support such models provide in estimating an ancestral state from an observation today. For the five models considered, this support varies in a predictable way depending on the type of model assumed. Moreover, the estimation of an ancestral state from the leaves of a phylogenetic tree exhibits further subtleties, along with a surprise: the independent estimates obtained from a star phylogeny may or may not be more informative than correlated estimates obtained from a binary tree, depending on the number of states, the size of the tree, and the substitution rate.

**Acknowledgments:** One of us (ES) presented this paper in 2012 at the University of Bordeaux, the London School of Economics, and the Institute for Mathematical Philosophy at the Ludwig Maximilian University in Munich, and received valuable comments. We are grateful for these. ES thanks the William F. Vilas Trust of the University of Wisconsin-Madison and MS thanks the NZ Marsden Fund and the Allan Wilson Centre for Molecular Ecology and Evolution.

**Appendix A**

**Proposition 1.1** Suppose that $X \to Y \to Z$ forms a Markov chain, where the state spaces for $X, Y$ and $Z$ are discrete, and where $\Pr(X = x \ \& \ Y = y) > 0$ and $\Pr(Y = y \ \& \ Z = z) > 0$ for all choices of states $x, y, z$ for $X, Y, Z$ respectively. Then the Data Processing Inequality is an equality if and only if $X, Y$ and $Z$ are mutually independent.

*Proof.* First, if $X, Y, Z$ are mutually independent then they are pairwise independent and so $I(X;Y) = I(X;Z) = I(Y;Z) = 0$ and thus equality holds trivially. Next, suppose that $\Pr(X = x \ \& \ Y = y) > 0$ and $\Pr(Y = y \ \& \ Z = z) > 0$ for all choices of states $x, y, z$ for $X, Y, Z$ respectively. Then $\Pr(X = x \ \& \ Y = y \ \& \ Z = z) > 0$ holds also (since $X \to Y \to Z$ is Markov chain). Suppose further that the DPI is an equality; we will show that $X, Y$ and $Z$ are independent. Since the DPI is an equality, $X \to Y \to Z$ and $X \to Z \to Y$ are both Markov chains (Cover and Thomas, 2006). We write $p(xyz)$ as short-hand for the probability $\Pr(X = x \ \& \ Y = y \ \& \ Z = z)$ and similarly for conditional and marginal probabilities (thus e.g. $p(x|z) = \Pr(X = x | Z = z)$). First observe that the positivity condition $p(xyz) > 0$ for all $(x, y, z)$ implies that $p(xy), p(xz), p(yz), p(x), p(y), p(z)$ are also strictly positive. Since $X \to Y \to Z$ is a Markov chain, and $p(xy) > 0$:

$$p(xyz) = p(z|xy)p(xy) = p(z|y)p(xy), \tag{1}$$

and since $X \to Z \to Y$ is also a Markov chain, and $p(xz) > 0$, we have: $p(xyz) = p(y|xz)p(xz) = p(y|z)p(xz)$. Applying Bayes' theorem the last term can be written as $\frac{p(z|y)p(y)}{p(z)} p(xz)$ (note that $p(z) > 0$) and so, combining this with Eqn. (1) gives:

$$p(xyz) = p(z|y)p(xy) = \frac{p(z|y)p(y)}{p(z)} p(xz),$$

and so:

$$p(z|y)p(xy)p(z) = p(z|y)p(y)p(xz). \tag{2}$$

Since $p(z|y) > 0$ (because $p(yz) > 0$) we can cancel this term on the left and right of Eqn. (2) to obtain:

$$p(xy)p(z) = p(y)p(xz). \tag{3}$$

Now, we can further write $p(xy) = p(x|y)p(y)$ and $p(xz) = p(x|z)p(z)$ which, upon substitution Eqn. (3) gives:

$$p(x|y)p(y)p(z) = p(y)p(x|z)p(z),$$

in other words: $p(x|y) = p(x|z)$ (noting that $p(y), p(z) > 0$). Now, this equation must hold for all choices of $x, y$ and $z$ so $p(x|y)$ must be constant as $y$ varies - which implies that $X$ and $Y$ are independent. Similarly $X$ and $Z$ are independent. Finally, reversing the two Markov chains gives that $Z$ is independent of $Y$. Thus $X, Y$ and $Z$ are pairwise independent. Moreover they are independent as a triple since $X \to Y \to Z$ is a Markov chain and so $p(xyz) = p(z|xy)p(xy) = p(z|y)p(x)p(y) = p(yz)p(x) = p(x)p(y)p(z)$.

## Appendix B: The $R_{ij}$ ratios with zero mutation at the infinite time limit

Consider the Moran model in population genetics, with two trait values $A$ and $B$ and population size $N$. Let $X_t \in \{0, 1, \ldots, N\}$ be the number of copies of $A$ in the population at time $t$. In this section we assume zero mutation, and we consider neutral evolution, selection for $A$, selection against $A$, and frequency-dependent selection (for the majority state and against the majority state). Since each of these Markov processes has absorbing states 0 and $N$ (because of zero mutation) eventually one allele will be fixed and the other lost. Let $E \in \{0, N\}$ be this end state, and $S \in \{0, 1, \ldots, N\}$ the starting state (thus $S = X_0$ and $E = \lim_{t \to \infty} X_t$). We are interested in comparing the ratio of conditional probabilities:

$$R_{ij} := \frac{\Pr(E = N | S = i)}{\Pr(E = N | S = j)},$$

for $i > j$ under the various models.

**Proposition 2.1**

**(i)** Under neutral evolution $R_{ij} = \dfrac{i}{j}$, for all $0 < i, j \leq N$

**(ii)** Under frequency-independent selection $R_{ij} = \dfrac{1-c^i}{1-c^j}$, for all $0 < i, j \leq N$, where $c$ is a positive constant with $c < 1$ when there is selection for $A$ and $c > 1$ when there is selection against $A$.

**(iii)** For any two values $i, j \in \{1, 2, \ldots, N\}$ with $i > j$ the $R_{ij}$ value for selection against $A$ exceeds the $R_{ij}$ value for neutral evolution, which in turn exceeds the $R_{ij}$ value for selection for $A$.

**(iv)** For frequency-dependent selection, where the fitness of trait $A$ is proportional to its frequency, the associated $R_{ij}$ value exceeds that for neutral evolution for all $i, j \in \{1, 2, \ldots, N\}$ with $i > j$ provided that $j < N/2$.

**(v)** For frequency-dependent selection, where the fitness of trait $A$ is proportional to the frequency of the alternative trait $B$, the associated $R_{ij}$ value is lower than that for neutral evolution for all $i, j \in \{1, 2, \ldots, N\}$ with $i > j$, provided that $j < N/2$.

**(vi)** In all the cases considered above we have $R_{ij} > 1$ for all $i > j$.

*Proof: Part (i):* For any integer $x: 0 \leq x \leq N$ it is a well-known result (for many neutral models) that

$$\Pr(E = N \mid S = x) = x/N$$

(see e.g. Eqn. (3.49) of Ewens (2010)), from which (i) immediately follows.

*Part (ii):* From Ewens (2010) Eqn. (3.66) we have, for any $x \in \{1, \ldots, N\}$:

$$\Pr(E = N \mid S = x) = \dfrac{1-c^x}{1-c^N},$$

for a positive constant $c$ which is greater than 1 for selection against $A$ and less than 1 for selection for $A$. Part (ii) now follows immediately.

*Part (iii).* By parts (i) and (ii), part (iii) is equivalent to the assertions that for $i > j$, if $c \in (0,1)$ then $\frac{1-c^i}{1-c^j} < \frac{i}{j}$, while if $i > j$ and $c > 1$ then $\frac{1-c^i}{1-c^j} > \frac{i}{j}$.

Now, using the identity $1-c^k = (1-c)(1+c+c^2+\cdots+c^{k-1})$ we have:

$$\frac{1-c^i}{1-c^j} = \frac{1+\cdots+c^{i-1}}{1+\cdots+c^{j-1}} \tag{4}$$

and so we wish to compare $\frac{1+\cdots+c^{i-1}}{1+\cdots+c^{j-1}}$ and $\frac{i}{j}$ which is equivalent to comparing:

$$\frac{1+\cdots+c^{i-1}}{i} \quad \text{and} \quad \frac{1+\cdots+c^{j-1}}{j}.$$

Now, the left-hand side is merely the average of the terms $c^k$ from $k=0$ up to $k=i-1$ while the right hand side is the average of these terms up to $j-1$ and since $i > j$ the left-hand side is smaller than the right when $c < 1$ and greater when $c > 1$. This completes the proof.

*Part (iv):* When selection is frequency-dependent we need to use the expression:

$$\Pr(E = N \mid S = i) = \frac{1 + \sum_{j=1}^{i-1} \prod_{k=1}^{j} \frac{g_k}{f_k}}{1 + \sum_{j=1}^{N-1} \prod_{k=1}^{j} \frac{g_k}{f_k}} \tag{5}$$

where $f_k$ and $g_k$ denote the fitnesses of the alleles A when B, respectively when $k$ individuals have allele type A (see e.g. Huang and Traulsen (2010), Eqn. (6)).

If we now take the fitness of trait $A$ to be proportional to its frequency, that is $f_k = a \cdot k$ for a constant $a > 0$, and take the fitness of trait $B$ to also be proportional to its frequency (with the same coefficient) -- that is, $g_k = a \cdot (N-k)$ -- then Eqn. (5) gives:

$$R_{ij} = \frac{\sum_{m=0}^{i-1} \binom{N-1}{m}}{\sum_{m=0}^{j-1} \binom{N-1}{m}}. \qquad (6)$$

As before, this ratio exceeds $\frac{i}{j}$ precisely when the average of the first $i$ terms $\binom{N-1}{m}$ (for $m = 0, 1, 2\ldots$) exceeds the average of the first $j$ terms $\binom{N-1}{m}$. And this holds for all $i > j$ with $j < N/2$.

*Part (v):* If we take the fitness of trait $A$ to be proportional to the frequency of $B$, that is $f_k = a \cdot (N-k)$ for a constant $a > 0$, and take the fitness of trait $B$ to be proportional to the frequency of $A$ (with the same coefficient), so $g_k = a \cdot k$, then, by Eqn. (5), we have:

$$R_{ij} = \frac{\sum_{m=0}^{i-1} \binom{N-1}{m}^{-1}}{\sum_{m=0}^{j-1} \binom{N-1}{m}^{-1}}, \qquad (7)$$

and this ratio is lower than $\frac{i}{j}$ precisely when the average of the first $i$ terms $\binom{N-1}{m}^{-1}$ (for $m = 0, 1, 2\ldots$) is lower than the average of the first $j$ terms $\binom{N-1}{m}^{-1}$. This holds for all $i > j$ with $j < N/2$.

*Part (vi):* The inequality $R_{ij} > 1$ for $i > j$ is trivial for neutral evolution, while for the other processes the inequality follows from Eqns. (5), (6) and (7), noting that when $i > j$ the $i$ terms in the numerator include the $j$ terms in the denominator along with some additional positive terms.

## Appendix C: The $R_{ij}$ ratios at finite time and nonzero mutation rate

The results in the previous section assume zero mutation and consider the infinite time limit. However they also have some bearing on what happens at finite time, and for non-zero mutation. First assume zero mutation, and consider the ratio $R_{ij}$ at finite times $t$. Since these ratios are continuous functions of $t$ and converge to values that satisfy the partial order described in Figure 3, this ordering also holds for $t$ sufficiently large (but finite). Select any such sufficiently value $t_0$ of $t$, and consider this ratio $R_{ij}$ at $t_0$ as a function of the mutation rates. Again, $R_{ij}$ is a continuous function of these mutation rates (with $t$ fixed to $t_0$) and so, for sufficiently small (but non-zero) mutation rates, the five $R_{ij}$ values will still be ordered as in Figure 3 at $t_0$.